\documentclass[preprint,aps,superscriptaddress,nofootinbib,tightenlines]{revtex4}
\usepackage{epsfig}
\usepackage{bm}

\def\OMIT#1{{}}

\newcommand{\beq}{\begin{equation}}
\newcommand{\eeq}{\end{equation}}
\newcommand{\bea}{\begin{eqnarray}}
\newcommand{\eea}{\end{eqnarray}}

\newcommand{\benn}{\begin{displaymath}}
\newcommand{\eenn}{\end{displaymath}}

\begin{document}

\begin{figure}[!t]
\vskip -1.5cm
\leftline{
{\epsfxsize=1.8in \epsfbox{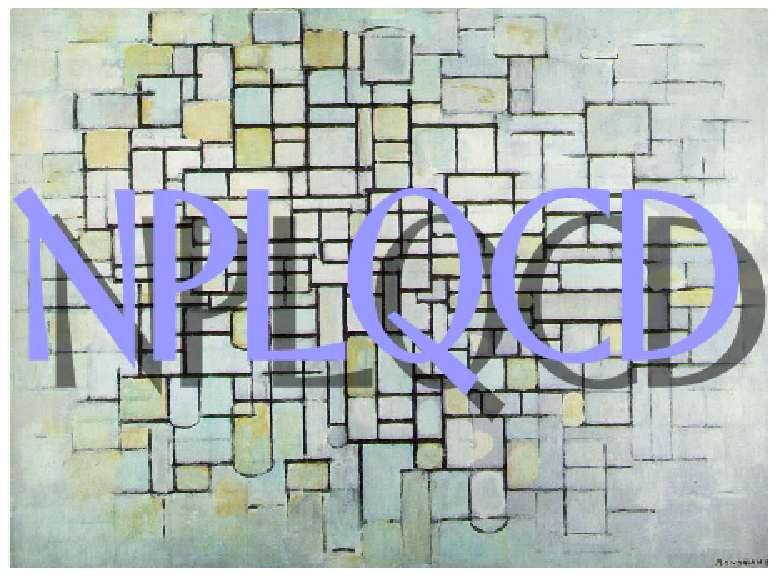}}}
\end{figure}

\preprint{\vbox{
\hbox{NT@UW-06-09}
\hbox{JLAB-THY-06-494}
\hbox{UNH-06-03}
}}

\vphantom{}
\title{\bf \LARGE 
Strong-Isospin Violation in the 
Neutron-Proton Mass Difference from Fully-Dynamical Lattice QCD and PQQCD}
\author{Silas R.~Beane}
\affiliation{Department of Physics, University of New Hampshire,
Durham, NH 03824-3568.}
\affiliation{Jefferson Laboratory, 12000 Jefferson Avenue, 
Newport News, VA 23606.}
\author{Kostas Orginos}
\affiliation{Jefferson Laboratory, 12000 Jefferson Avenue, 
Newport News, VA 23606.}
\affiliation{Department of Physics, College of William and Mary, Williamsburg,
  VA 23187-8795.}
\author{Martin J.~Savage}
\affiliation{Department of Physics, University of Washington, 
Seattle, WA 98195-1560.\\
\qquad}
\collaboration{ NPLQCD Collaboration }
\noaffiliation
\vphantom{}
\vskip 0.8cm
\begin{abstract} 
\vskip 0.5cm
\noindent We determine the strong-isospin violating component of the
neutron-proton mass difference from fully-dynamical lattice QCD and
partially-quenched QCD calculations of the nucleon mass, constrained
by partially-quenched chiral perturbation theory at one-loop level.
The lattice calculations were performed with domain-wall valence
quarks on MILC lattices with rooted staggered sea-quarks at a lattice
spacing of $b=0.125~{\rm fm}$, lattice spatial size of $L=2.5~{\rm
fm}$ and pion masses ranging from $m_\pi\sim 290~{\rm MeV}$ to $\sim
350~{\rm MeV}$.  At the physical value of the pion mass, we predict
$M_n-M_p\big|^{d-u} = 2.26\pm 0.57 \pm 0.42 \pm 0.10~{\rm MeV}$ where
the first error is statistical, the second error is due to the
uncertainty in the ratio of light-quark masses, $\eta=m_u/m_d$,
determined by MILC~\cite{Aubin:2004fs}, and the third error is an
estimate of the systematic due to chiral extrapolation.
\end{abstract}
\maketitle

\vfill\eject

\section{Introduction}

\noindent It is a basic property of our universe that the neutron is
slightly more massive than the proton.  The electroweak interactions
are responsible for this mass difference, which receives contributions
from two sources.  The strong isospin breaking contribution (also
known as charge-symmetry breaking, for a review see
Ref.~\cite{Miller:2006tv}) is due to the difference in the masses of
the up and down quarks, ultimately determined by the values of the
Yukawa couplings in the Standard Model of electroweak interactions and
the vacuum expectation value of the Higgs field. The other
contribution arises from the fact that the proton and neutron carry
different electromagnetic charges.  The experimental neutron-proton
mass difference of $M_n-M_p = 1.2933317\pm 0.0000005~{\rm
MeV}$~\cite{Eidelman:2004wy} receives an estimated electromagnetic
contribution of~\cite{Gasser:1982ap} $M_n-M_p\big|^{\rm em} = -0.76\pm
0.30~{\rm MeV}$, and the remaining mass difference is due to a strong
isospin breaking contribution of $M_n-M_p\big|^{d-u}= 2.05\mp
0.30~{\rm MeV}$.

The recent suggestion of a vast landscape of possible universes
emerging from string theory~\cite{Kachru:2003aw,Susskind:2003kw}, and
the previously hypothesized multi-universe scenario in general, have
led to a resurgence of interest in the anthropic principle, and how it
may provide a useful way of constraining fundamental parameters of
nature. For this reason and others it is of interest to understand how
quantities that affect nuclear physics and the production of elements
depend upon the fundamental parameters: the length scale of the strong
interactions, $\Lambda_{\rm QCD}$, the light-quark masses, $m_u$,
$m_d$ and $m_s$, and the electromagnetic coupling, $\alpha_e$.  For
instance, the basic fact that the neutron-proton mass difference is
larger than the electron mass is central to the evolution and content
of our universe (for a recent detailed discussion of the impact of an
unstable proton see Ref.~\cite{Hogan:2006xa}.).  While a universe with
a stable neutron would have a rich periodic table of nuclei, it is far
from clear that the organic chemistry required for carbon-based life
would exist~\cite{Hogan:2006xa,Agrawal:1997gf,Agrawal:1998xa}.

In this work we have performed lattice calculations of the nucleon
mass with domain-wall valence quark masses tuned to the staggered
sea-quark masses of the MILC configurations (a mixed-action ``QCD''
calculation), and with valence quark masses that differ from the
sea-quark masses (a mixed-action partially-quenched, ``PQ'',
calculation).  As the computer resources do not presently exist to
perform such calculations at the physical values of the light-quark
masses, they are performed at QCD pion masses of $m_\pi^{\rm latt.}
\sim 290~{\rm MeV}$ and $m_\pi^{\rm latt.} \sim 350~{\rm MeV}$ and a
range of partially-quenched pion masses in between.  These results are
combined with the formal framework of heavy-baryon chiral perturbation
theory (HB$\chi$PT~\cite{Jenkins:1990jv,Jenkins:1991ne}) and
partially-quenched heavy-baryon chiral perturbation theory
(PQHB$\chi$PT~\cite{Labrenz:1996jy,Savage:2001dy,Chen:2001yi,Beane:2002vq}), and the
recent precise lattice extraction of the light-quark mass ratio,
$m_u/m_d = 0.43\pm 0.01\pm 0.08$~\cite{Aubin:2004fs} by the MILC
collaboration, to calculate the strong isospin-breaking
contribution to the neutron-proton mass-difference.

\section{The Formal Framework}

\noindent Nucleon observables have a systematic loop expansion about
the limit of vanishing quark masses and external momenta defined by
HB$\chi$PT~\cite{Jenkins:1990jv,Jenkins:1991ne}.  Extensive
phenomenology has been performed with HB$\chi$PT, where it is found
that some observables appear to converge quite rapidly in the chiral
expansion, while others are less convergent. This state of affairs is
magnified when dealing with three light flavors rather than two.
Recently, the heavy-baryon formalism has been extended to describe the
situation where the sea-quark masses differ from the valence quark
masses~\cite{Labrenz:1996jy,Savage:2001dy,Chen:2001yi,Beane:2002vq}, as is the case
in partially-quenched lattice calculations.

The mass of the proton, which has valence quark content $uud$, when
computed on configurations with sea-quarks, $j$ and $l$, of mass $m_j$
and $m_l$ has the form~\cite{Beane:2002vq} (we have used the opposite
sign convention for the mass counterterms)
\begin{eqnarray}
& & M_p \ = \
 M_0
\ +\ (m_u+m_d)\left(\alpha+\beta\right)
\ +\ 
2\sigma \left(m_j+m_l\right)
\ +\ {1\over 3}\left(2 \alpha-\beta\right)(m_u-m_d)
\nonumber\\
& & \qquad
-  \ 
{1\over 8\pi f^2}\left(\ 
{g_A^2\over 3}\left[\ 
m_{uu}^3+m_{ud}^3+2 m_{ju}^3+ 2 m_{lu}^3 + 3 G_{\eta_u , \eta_u}
\ \right]
\right.\nonumber\\
& & \left. \qquad
\ \ +\ {g_1^2\over 12}\left[\ 
m_{uu}^3 - 5 m_{ud}^3 + 
3 m_{jd}^3 + 2 m_{ju}^3 + 3 m_{ld}^3 
+ 2 m_{lu}^3 + 3 G_{\eta_u , \eta_u}
+ 6 G_{\eta_u , \eta_d} + 3 G_{\eta_d , \eta_d}
\  \right]
\right.\nonumber\\
& & \left. \qquad
\ \ +\ {g_A g_1\over 3}\left[\ 
m_{ju}^3 + m_{lu}^3 - m_{ud}^3 + 2 m_{uu}^3 + 3  G_{\eta_u , \eta_d}
+ 3 G_{\eta_u , \eta_u}
\  \right]
\right.\nonumber\\
& & \left. \qquad
\ \ + \ {g_{\Delta N}^2\over 9\pi}\left[\ 
5 F_{ud} + F_{uu} + F_{ju} + F_{lu} + 2 F_{jd} + 2 F_{ld}
+ 2 E_{\eta_d , \eta_d} +2 E_{\eta_u , \eta_u} 
- 4 E_{\eta_u , \eta_d}
\ \right]\right)
,
\label{eq:protmass}
\end{eqnarray}
where $\alpha,\beta,\sigma$ are the counterterms that enter at ${\cal
O}(m_q)$, $g_A$ is the nucleon axial coupling constant, $g_{\Delta
N}$ is the $\Delta$-nucleon coupling constant, and $g_1$ is
the nucleon coupling to the flavor-singlet meson field, which makes no
contribution in the QCD limit where $m_j\rightarrow m_u$ and
$m_l\rightarrow m_d$.  The notation is such that $m_{ab}$ is the mass
of the pseudo-Goldstone boson composed of quarks $a$ and $b$.  The
functions $G_{\eta_a , \eta_b}$, and $E_{\eta_a , \eta_b}$ are
$G_{\eta_a ,\eta_b}= {\cal
H}_{\eta_a\eta_b}(m_{\eta_a}^{3},m_{\eta_b}^{3},m_X^{3})$ and
$E_{\eta_a , \eta_b}={\cal
H}_{\eta_a\eta_b}(F_{\eta_a},F_{\eta_b},F_{X})$, respectively, where
the function ${\cal H}_{\eta_a\eta_b}$ is given by
\begin{eqnarray}
{\cal H}_{ab}( A, B, C) & = & 
-{1\over 2}\left[\ 
{(m_{jj}^2-m_{\eta_a}^2)(m_{ll}^2-m_{\eta_a}^2)\over 
(m_{\eta_a}^2-m_{\eta_b}^2)(m_{\eta_a}^2-m_X^2)}\  A
-
{(m_{jj}^2-m_{\eta_b}^2)(m_{ll}^2-m_{\eta_b}^2)\over 
(m_{\eta_a}^2-m_{\eta_b}^2)(m_{\eta_b}^2-m_X^2)}\  B
\right.
\nonumber\\ 
& & \left.\qquad 
\ +\ 
{(m_X^2-m_{jj}^2)(m_X^2-m_{ll}^2)\over 
(m_X^2-m_{\eta_a}^2)(m_X^2-m_{\eta_b}^2)}\  C
\ \right]
\ \ \ ,
\label{eq:HPsdef}
\end{eqnarray}
and where the mass, $m_X$, is given by $m_X^2 = {1\over 2}\left(
m_{jj}^2+ m_{ll}^2 \right)$.  The $\Delta$-loop function (e.g. $F_\pi
= F(m_\pi,\Delta,\mu) $) is given by
\begin{eqnarray}
 F(m,\Delta,\mu) & = & 
\left(m^2-\Delta^2\right)\left(
\sqrt{\Delta^2-m^2} \log\left({\Delta -\sqrt{\Delta^2-m^2+i\epsilon}\over
\Delta +\sqrt{\Delta^2-m^2+i\epsilon}}\right)
-\Delta \log\left({m^2\over\mu^2}\right)\ \right)
\nonumber\\
& - & 
{1\over 2} m^2 \Delta  \log\left({m^2\over\mu^2}\right)
\ \ \ .
\label{eq:massfun}
\end{eqnarray}
The nucleon mass has been computed to one higher order in the
partially-quenched chiral expansion, ${\cal O}(m_q^2)$, in
Ref.~\cite{Tiburzi:2005na}, but with the small number of different
quark masses available to us in our lattice calculation we will be
unable to make use of that work.  In the QCD limit, the proton mass
computed from eq.~(\ref{eq:protmass}) becomes the QCD proton mass
computed previously with HB$\chi$PT,
\begin{eqnarray}
M_p & = & M_0 +  \left(\alpha+\beta+2\sigma\right)(m_u+m_d)
\ +\ {1\over 3}\left(2 \alpha-\beta\right)(m_u-m_d)
\nonumber\\
& - &
{1\over 8\pi f^2}\left[\ {3\over 2} g_A^2 m_\pi^3
\ +\ {4 g_{\Delta N}^2\over 3\pi} F_\pi\ \right]
\ \ \ ,
\label{eq:protQCD}
\end{eqnarray}
as required.  The expansion of the neutron mass can be recovered from
the expansion of the proton mass by interchanging the up and down
quark masses, $u\leftrightarrow d$.  At the order to which we are
working it is most convenient to replace the explicit quark masses in
the expression for the proton mass with the leading-order expression
for the pion mass to yield
\begin{eqnarray}
M_p & = & M_0 + \left(\overline{\alpha} + \overline{\beta} + 2 \overline{\sigma}\right) m_\pi^2 
\ -\ 
{1\over 3}\left(2 \overline{\alpha} - \overline{\beta}\right)\left({1 - \eta\over 1+\eta}\right)\ 
m_\pi^2\  
\nonumber\\
& - &
{1\over 8\pi f^2}\left[\ {3\over 2} g_A^2 m_\pi^3
\ +\ {4 g_{\Delta N}^2\over 3\pi} F_\pi\ \right]
\ \ \ ,
\label{eq:protmasspion}
\end{eqnarray}
where $\eta=m_u/m_d$.
The neutron mass is recovered by making the replacement $\eta\rightarrow
1/\eta$, and consequently
\begin{eqnarray}
M_n \ -\  M_p\big|^{d-u} & = & 
  {2\over 3}\left(2 \overline{\alpha} - \overline{\beta}\right)
\left({1 - \eta\over 1+\eta}\right)\ 
m_\pi^2\
\ \ \ .
\label{eq:mnminusmp}
\end{eqnarray}
The one-loop contributions at ${\cal O}(m_q^{3/2})$ cancel in the mass-difference,
as the pions are degenerate up to ${\cal O}(m_q^2)$.
Analogous expressions for the partially-quenched proton masses can be found
in Appendix~\ref{app:pqmass}. 

The practical message one should take from the functional form of the proton mass in
eq.~(\ref{eq:protmass}) is that partial-quenching allows an extraction of
isospin-violating quantities from isospin-symmetric lattices.

\section{Details of the Lattice Calculation and Analysis }

\begin{figure}[!ht]
\centerline{{\epsfxsize=4.5in \epsfbox{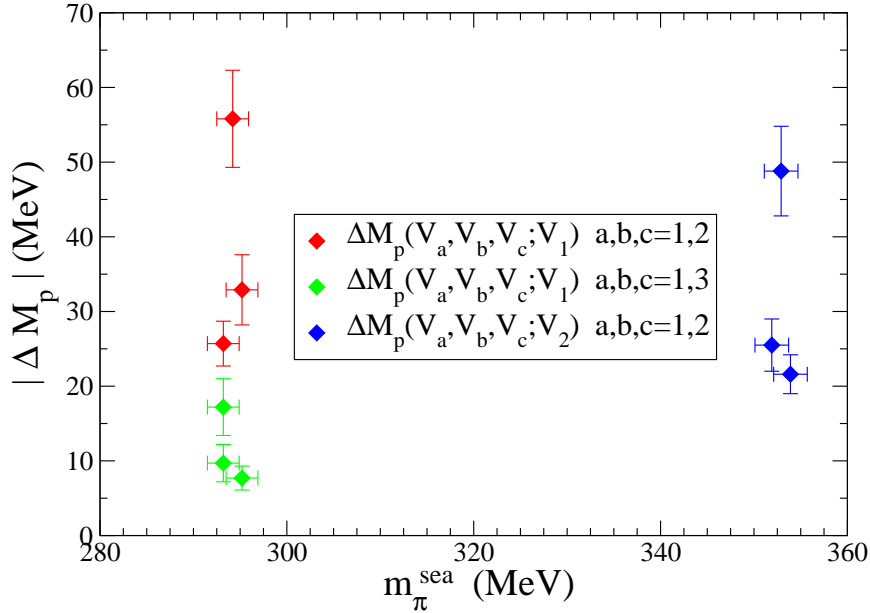}}} 
\vskip 0.15in
\noindent
\caption{\it 
The partially-quenched proton mass differences (in MeV) calculated from the $b m_l=0.007$ and
$0.010$ MILC lattices plotted vs the pion mass composed of sea quarks.
Various data have been displaced 
horizontally by small amounts for display purposes. A lattice spacing of $b=0.125~{\rm fm}$ has been used.
}
\label{fig:pqdata}
\vskip .2in
\end{figure}

\noindent 
Our computation uses the mixed-action lattice QCD scheme developed by
LHPC~\cite{Renner:2004ck,Edwards:2005kw} using domain-wall valence
quarks from a smeared-source on $N_f=2+1$
asqtad-improved~\cite{Orginos:1999cr,Orginos:1998ue} MILC
configurations generated with rooted~\footnote{For recent discussions
of the ``legality'' of the mixed-action and rooting procedures, see
Ref.~\cite{Durr:2004ta,Creutz:2006ys,Bernard:2006vv,Durr:2006ze,Hasenfratz:2006nw}.}
staggered sea quarks~\cite{Bernard:2001av} that are hypercubic-smeared
(HYP-smeared)~\cite{Hasenfratz:2001hp,DeGrand:2002vu,DeGrand:2003in,Durr:2004as}. In
the generation of the MILC configurations, the strange-quark mass was
fixed near its physical value, $b m_s = 0.050$, (where $b=0.125~{\rm
fm}$ is the lattice spacing) determined by the mass of hadrons
containing strange quarks.  
The two light quarks in the configurations
are degenerate (isospin-symmetric).  The domain-wall height is $m=1.7$
and the extent of the extra dimension is $L_5=16$.  The MILC lattices
were ``chopped'' using a Dirichlet boundary condition from 64 to 32
time-slices to save time in propagator generation.  In order to
extract the terms in the mass expansion, we computed a number of sets
of propagators corresponding to different valence quark masses, as
shown in Table~I.
%
\begin{table}[ht]
\vskip 0.1in
\begin{tabular}{|c|c|c|c|c|c|c|}
\hline
\label{table:configs}
 Ensemble        
& \quad Theory \quad 
&\quad  $b m_l$ \quad
&\  $b m_s$ \   
&\quad $b m_{dwf}$\quad  
&\quad $10^3 \times b m_{res}$\quad 
& \# props.  \\
       \hline 
\ 2064f21b679m007m050 \ & \,QCD&  0.007 ($V_1$) & 0.050   & 0.0081 ($V_1$)     & $1.604\pm 0.038$        & 468$\times$3 \\
2064f21b679m007m050 & PQQCD& 0.007 ($V_1$)& 0.050   & 0.0138 ($V_2$)    & $1.604\pm 0.038$        & 367$\times$3 \\
2064f21b679m007m050 & PQQCD& 0.007 ($V_1$)& 0.050   & 0.0100 ($V_3$)     & $1.604\pm 0.038$        & 367$\times$2 \\
2064f21b679m010m050 & \,QCD& 0.010 ($V_2$)& 0.050   & 0.0138 ($V_2$)    & $1.552\pm 0.027$       & 658$\times$3 \\
2064f21b679m010m050 & PQQCD& 0.010 ($V_2$)& 0.050   & 0.0081 ($V_1$)    & $1.552\pm 0.027$       & 658$\times$1 \\
  \hline
  \end{tabular}
  \caption{The parameters of the MILC gauge configurations and 
domain-wall propagators used in this work. 
For each propagator the extent of the fifth dimension is $L_5=16$.
The notation of quarks, $V_1, V_2, V_3$, is defined in the text.
The last column is the number of propagators generated, and corresponds to the
number of lattices times the number of different locations of sources on each lattice.
}
\end{table}
On 468 $b m_l=0.007$ (denoted by $V_1$) lattices we have computed
three sets corresponding to the QCD point with a valence-quark mass of
$b m_{dwf}=0.0081$ ($V_1$), three sets on 367 $b m_l=0.007$ lattices
with a valence quark mass of $b m_{dwf}=0.0138$ (denoted by $V_2$),
and two sets with a valence quark mass $b m_{dwf}=0.0100$ (denoted by
$V_3$). On 658 of the $b m_l=0.010$ ($V_2$) lattices we have computed
three sets at the QCD point with a valence-quark mass of $b
m_{dwf}=0.0138$ ($V_2$) and one set with a valence quark mass of $b
m_{dwf}=0.0081$ ($V_1$). The parameters used to generate the QCD-point
light-quark propagators have been ``matched'' to those used to
generate the MILC configurations so that the mass of the pion computed
with the domain-wall propagators is equal (to few-percent precision)
to that of the lightest staggered pion computed with the same
parameters as the gauge configurations~\cite{Bernard:2001av}.  The
lattice calculations were performed with the {\it Chroma} software
suite~\cite{Edwards:2004sx,sse2} on the high-performance computing
systems at the Jefferson Laboratory (JLab).
\begin{figure}[!ht]
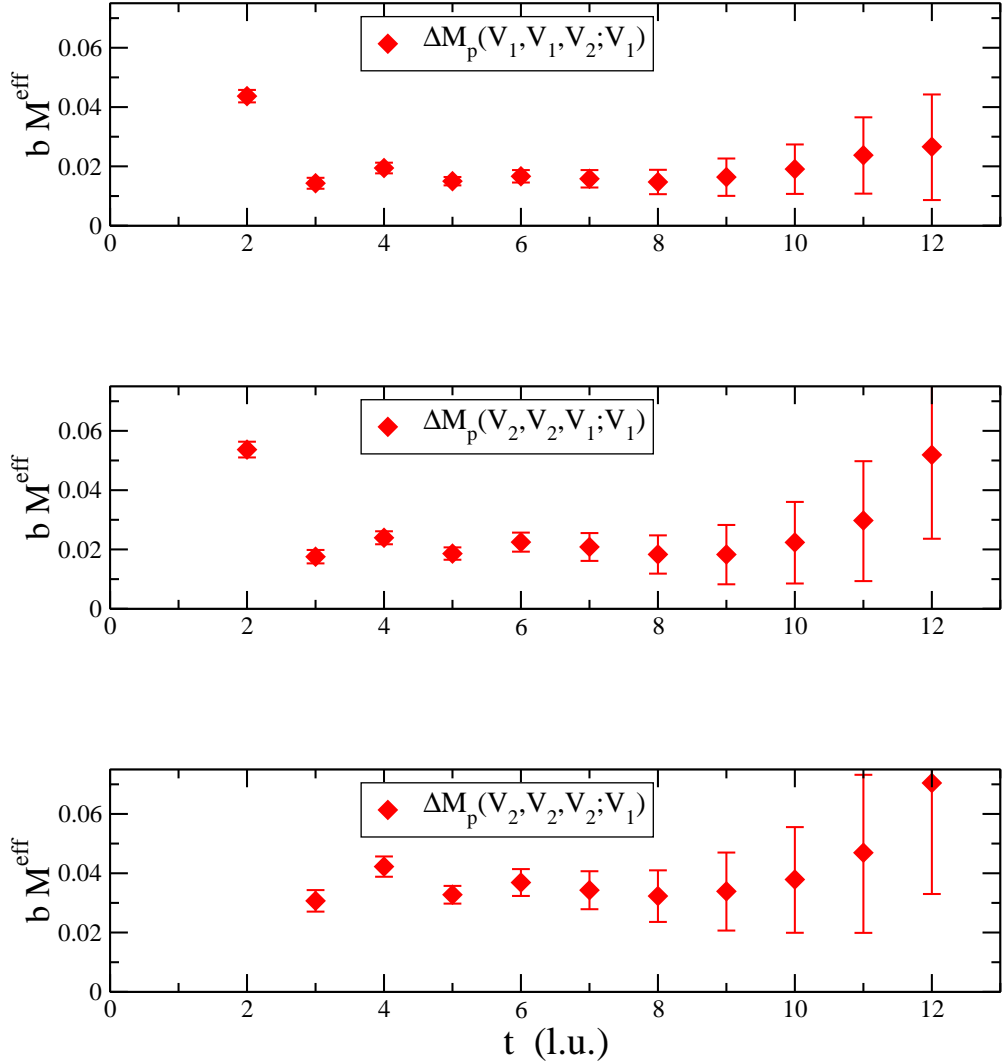

\vskip 0.47in
\centerline{{\epsfxsize=5.2in \epsfbox{Effmass_v1v1v2v1_part2.eps}}}
\vskip 0.44in
\centerline{{\epsfxsize=5.2in \epsfbox{Effmass_v2v2v1v1_part2.eps}}}
\vskip 0.44in
\centerline{{\epsfxsize=5.2in \epsfbox{Effmass_v2v2v2v1_part2.eps}}}
\noindent
\caption{\it Effective mass plots for the ratios of proton correlation functions
that give rise to the mass splittings $\Delta M_p(V_a,V_b,V_c;V_1)$ with $a,b,c$=$1,2$.}
\label{fig:DELTAMpeffmassA}
\end{figure}
\begin{figure}[!ht]
\vskip 0.44in
\centerline{{\epsfxsize=5.2in \epsfbox{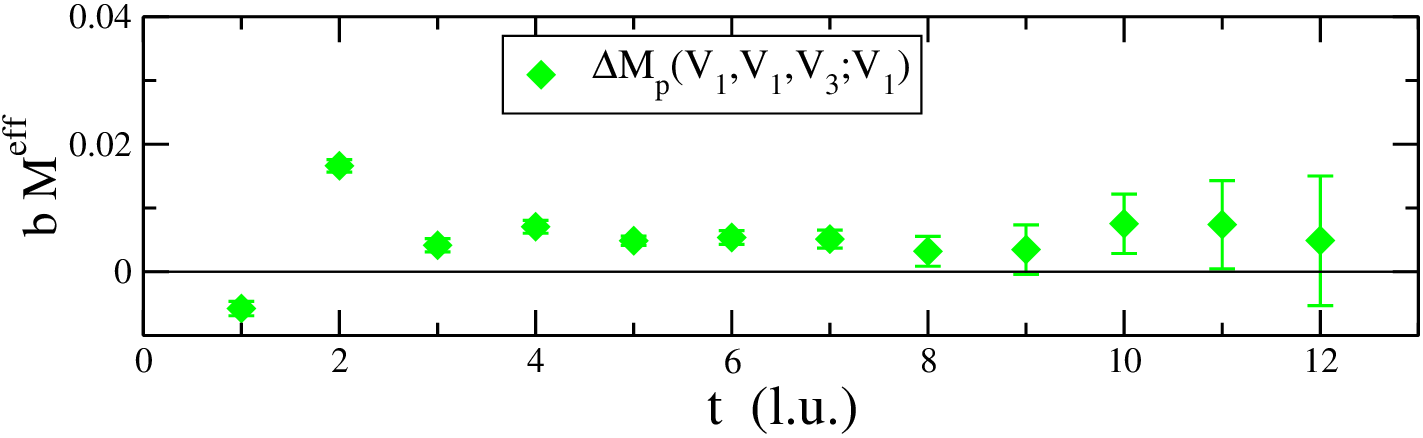}}}
\vskip 0.43in
\centerline{{\epsfxsize=5.2in \epsfbox{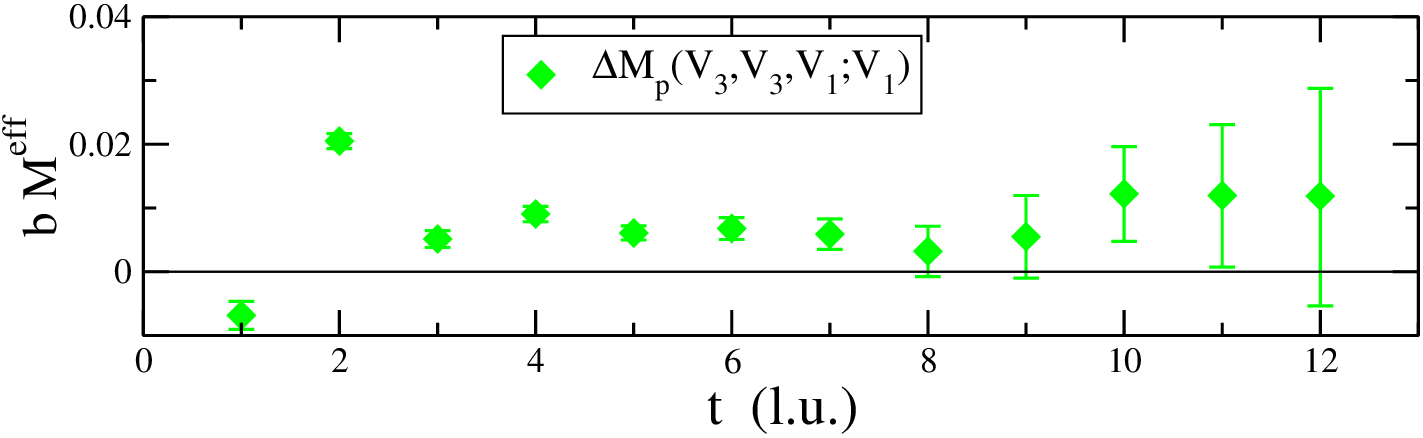}}}
\vskip 0.43in
\centerline{{\epsfxsize=5.2in \epsfbox{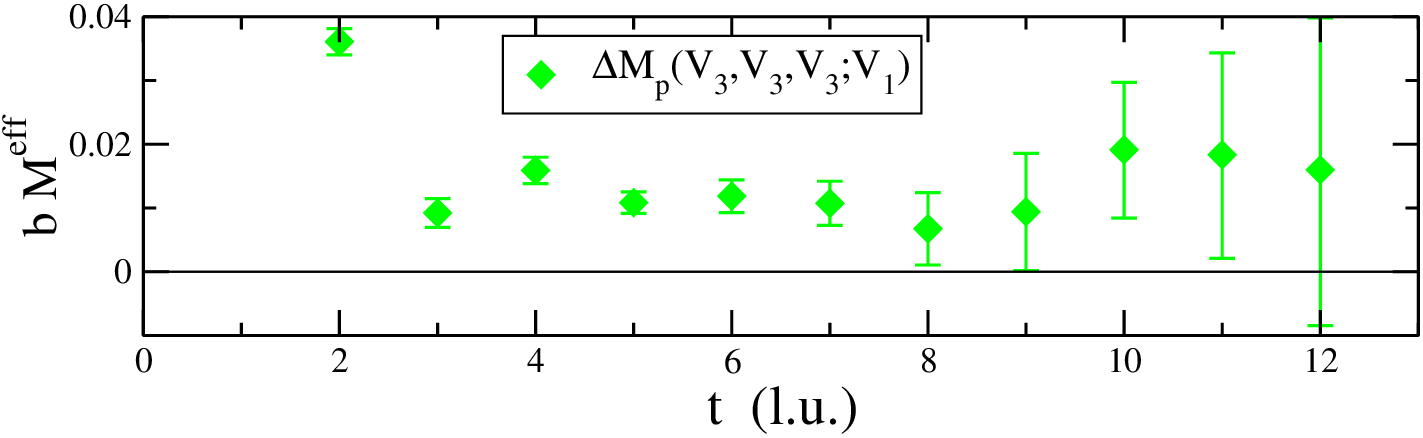}}}
\noindent
\caption{\it Effective mass plots for the ratios of proton correlation functions
that give rise to the mass splittings $\Delta M_p(V_a,V_b,V_c;V_1)$with $a,b,c$=$1,3$.}
\label{fig:DELTAMpeffmassB}
\end{figure}
\begin{figure}[!ht]
\vskip 0.44in
\centerline{{\epsfxsize=5.2in \epsfbox{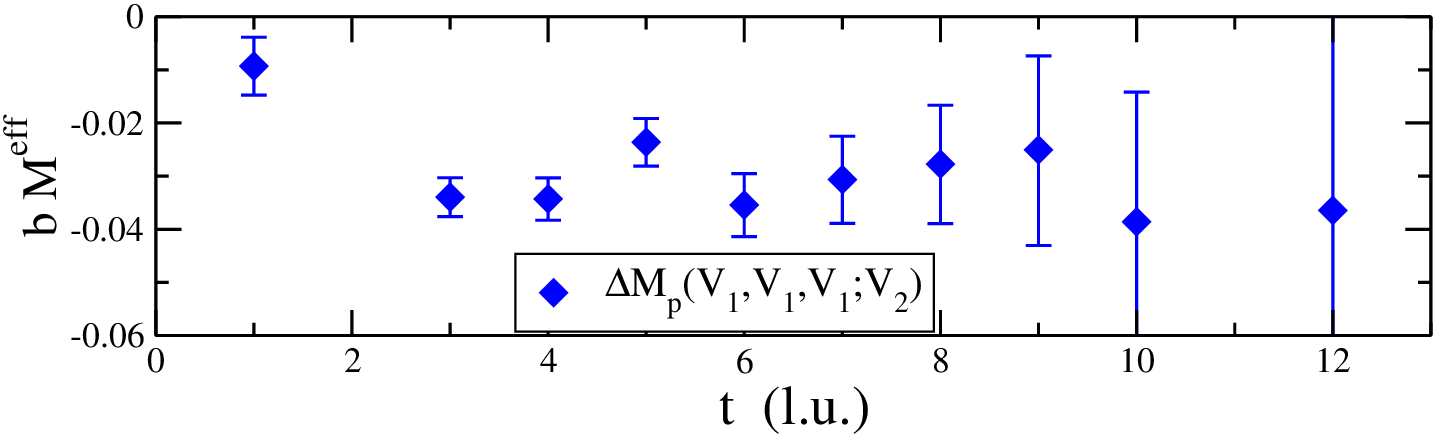}}}
\vskip 0.43in
\centerline{{\epsfxsize=5.2in \epsfbox{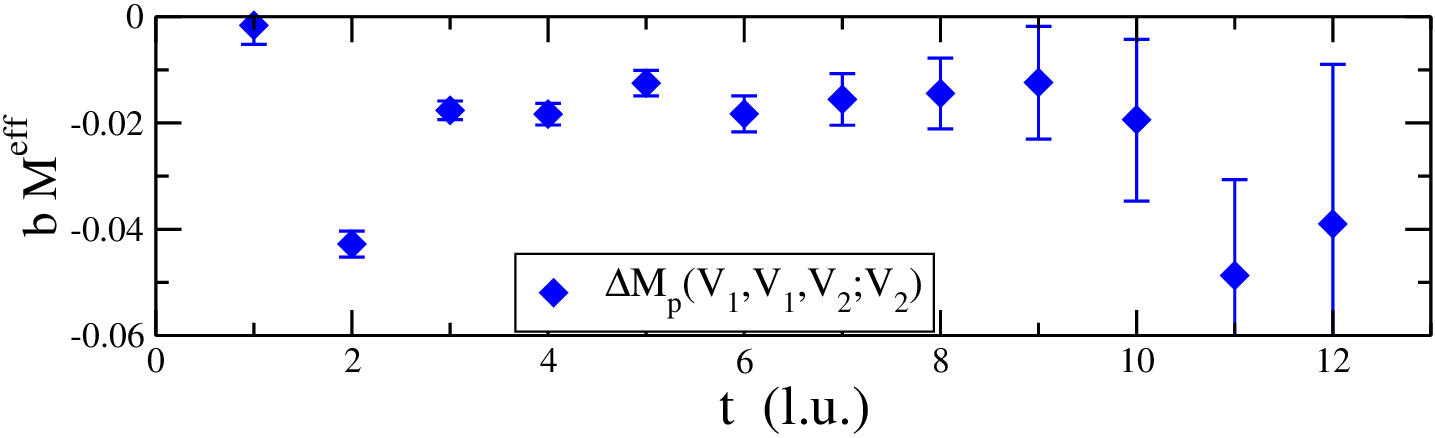}}}
\vskip 0.43in
\centerline{{\epsfxsize=5.2in \epsfbox{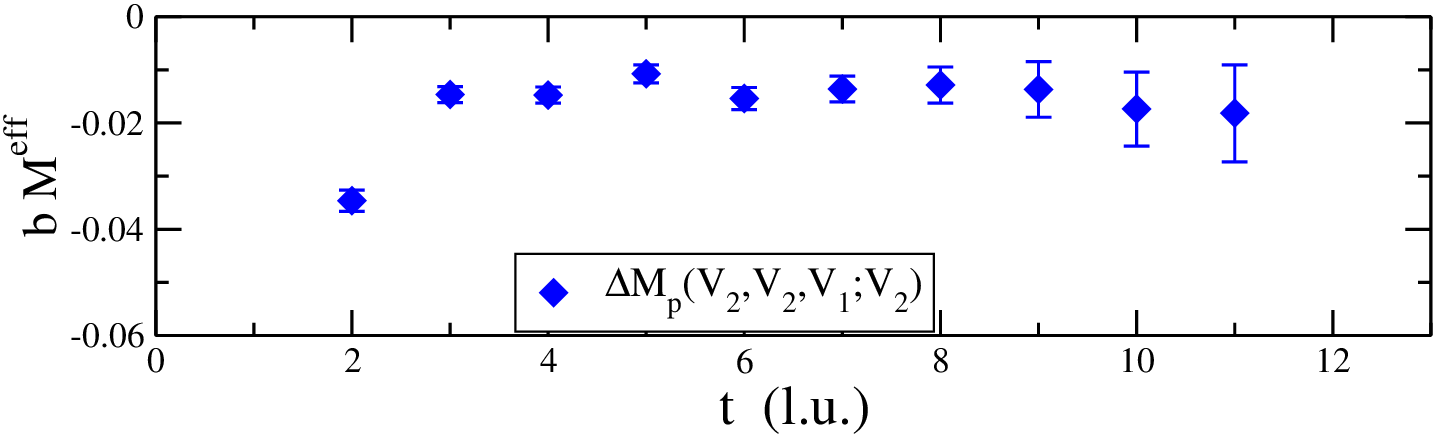}}}
\noindent
\caption{\it Effective mass plots for the ratios of proton correlation functions
that give rise to the mass splittings $\Delta M_p(V_a,V_b,V_c;V_2)$ with $a,b,c$=$1,2$.}
\label{fig:DELTAMpeffmassC}
\end{figure}
Various proton and pion correlation functions were constructed from
the three (distinct-mass) light-quark propagators that were generated
on the $b m_l=0.007$ lattices, and the two light-quark propagators
that were generated on the $b m_l=0.010$ lattices. Differences
between the various proton states were also constructed.  It is useful to
define the proton mass splitting:
\begin{eqnarray}
\Delta M_p(V_a,V_b,V_c;V_d)\ \equiv \  M_p(V_a,V_b,V_c;V_d) - M_p(V_d,V_d,V_d;V_d) \ ,
\label{eq:deltaprotmass}
\end{eqnarray}
where the indices $a,b,c$ range over $1,2,3$. Effective mass plots for
the ratios of proton correlators that give rise to these mass
splittings are displayed in figs.~\ref{fig:DELTAMpeffmassA},
\ref{fig:DELTAMpeffmassB} and ~\ref{fig:DELTAMpeffmassC}.  Results for
the extracted proton mass splittings are given in Table~\ref{table:pq}
and displayed in fig.~\ref{fig:pqdata}. The various pion masses
relevant to the analysis are also shown in Table~\ref{table:pq}. Using
\begin{table}[t]
\begin{tabular}{|c|c|c|c|}
\hline
 Quantity &\    Mass (Difference) (l.u.)\   &\  Mass (Difference) (MeV)\  &\ Fitting Range\ \\
\cline{1-4}
\cline{1-4}
$m_\pi(V_1,V_1;V_1)$  & $0.1864 \pm 0.0011$  & $294.2 \pm 1.7  $  & 
$ 5\rightarrow 15$ \\
\cline{1-4} 
$m_\pi(V_1,V_2;V_1)$ & $0.2066 \pm 0.0010$  & $ 326.2 \pm 1.6 $   & $ 5\rightarrow 15$ \\
\cline{1-4} 
$m_\pi(V_2,V_2;V_1)$ & $0.22473 \pm 0.00091$  & $ 354.4 \pm 1.4 $   & $
5\rightarrow 15$  \\
\cline{1-4} 
$m_\pi(V_1,V_3;V_1)$ & $0.1929 \pm 0.0012$  & $ 304.5 \pm 1.9 $   & $
5\rightarrow 15$  \\
\cline{1-4} 
$m_\pi(V_3,V_3;V_1)$ & $0.1996 \pm 0.0011$  & $ 315.1 \pm 1.8 $   & $
5\rightarrow 15$  \\
\cline{1-4} 
$m_\pi(V_1,V_1;V_2)$ & $0.1844 \pm 0.0013$  & $ 291.0  \pm 2.1 $   & $
5\rightarrow 15$  \\
\cline{1-4} 
$m_\pi(V_1,V_2;V_2)$ & $0.2050 \pm 0.0012$  & $ 323.7  \pm 1.0 $   & $
5\rightarrow 15$  \\
\cline{1-4} 
$m_\pi(V_2,V_2;V_2)$  & $0.2236 \pm 0.0011$  & $ 352.9  \pm 1.8 $   & $
5\rightarrow 15$  \\
\hline\hline
\ $\Delta M_p(V_1,V_1,V_2;V_1)$ \ & $0.0163\pm  0.0019 $  & $ 25.7 \pm 3.0   $
& $ 5\rightarrow 12$ \\
\cline{1-4} 
\ $\Delta M_p(V_2,V_2,V_1;V_1)$ \ & $0.0209 \pm 0.0029 $  & $32.9 \pm 4.7   $
& $ 5\rightarrow 12$ \\
\cline{1-4} 
\ $\Delta M_p(V_2,V_2,V_2;V_1)$ \ & $0.0353 \pm 0.0041 $  & $ 55.8 \pm 6.5  $
& $ 5\rightarrow 12$ \\
\cline{1-4} 
\ $\Delta M_p(V_1,V_1,V_3;V_1)$ \ & $0.0049\pm  0.0010 $  & $ 7.7 \pm 1.6   $
& $ 5\rightarrow 11$\\
\cline{1-4} 
\ $\Delta M_p(V_3,V_3,V_1;V_1)$ \ & $0.0061 \pm 0.0016 $  & $9.7 \pm 2.5   $
&  $ 5\rightarrow 11$ \\
\cline{1-4} 
\ $\Delta M_p(V_3,V_3,V_3;V_1)$ \ & $ 0.0109 \pm 0.0024 $  & $17.2 \pm 3.8   $
&  $ 5\rightarrow 11$ \\
\cline{1-4} 
\ $\Delta M_p(V_1,V_1,V_1;V_2)$ \ & $-0.0309\pm  0.0038 $  & $ -48.8 \pm 6.0
$   &  $ 4\rightarrow 11$ \\
\cline{1-4} 
\ $\Delta M_p(V_1,V_1,V_2;V_2)$ \ & $-0.0161\pm  0.0022 $  & $ -25.5 \pm 3.5
$   &  $ 4\rightarrow 11$ \\
\cline{1-4} 
\ $\Delta M_p(V_2,V_2,V_1;V_2)$ \ & $-0.0137\pm  0.0016 $  & $ -21.6 \pm 2.6
$   &  $ 5\rightarrow 12$ \\
\cline{1-4} 
\end{tabular}
\caption{The pion masses and proton  mass differences calculated on the 
$b m_l=0.007$ and $b m_l=0.010$  MILC lattices.
The notation of valence and sea quarks, $V_{1,2,3}$, is defined in the text.
A lattice spacing of $b=0.125~{\rm fm}$ has been used.
}
\label{table:pq}
\end{table}
this data in conjunction with the PQHB$\chi$PT formulas given above
and in the appendix, the coefficients $\overline{\alpha}$,
$\overline{\beta}$, $g_{\Delta N}$, and $g_1$ that appear in
eq.~(\ref{eq:protmasspion}) were extracted using various nonlinear
fitting techniques.
A tree-level, ${\cal O}(m_q)$, analysis of the mass
differences shown in Table~\ref{table:pq} using the expressions for
the partially-quenched proton masses given in the appendix, allows for
an extraction of the isospin-breaking coefficient
$\left(2\overline{\alpha}-\overline{\beta}\right)/3$, and the
isospin-conserving coefficient $ \overline{\alpha}+\overline{\beta}$,
as shown in Table~\ref{table:params}.  This leads to a prediction for
the strong isospin breaking at the physical quark masses, at ${\cal
O}(m_q)$, of
\begin{eqnarray}
M_n-M_p\big|^{d-u} & = & 1.96\pm 0.92 \pm 0.37~{\rm MeV}
\ \ \ ,
\label{eq:treelevelnumbers}
\end{eqnarray}
as shown in Table~\ref{table:properties}.
The first error is statistical and the second error is due to the 
uncertainty in the determination of the ratio of light quark masses,
$\eta=m_u/m_d = 0.43\pm 0.01\pm 0.08$
by MILC~\cite{Aubin:2004fs}.

The analysis of the partially-quenched data at ${\cal O}(m_q^{3/2})$
introduces contributions to the various proton masses that vanish in
the QCD limit, a characteristic feature of the partially-quenched
theory.  In this case there is a contribution from loop diagrams
involving the flavor-singlet field --- due to a mismatch between the
ordinary flavor singlet and that of the graded-group --- which depends
on an axial coupling constant, $g_1$~\cite{Beane:2002vq}, that must
also be determined from the lattice data.  Fortunately, with multiple
partially-quenched proton states, $g_1$ can be extracted
simultaneously with $\left(2\overline{\alpha}-\overline{\beta}\right)/3$, 
$\overline{\alpha}+\overline{\beta}$, and $g_{\Delta N}$.
However, given that $\overline{\alpha}$ and $\overline{\beta}$ enter at
tree-level, while $g_1$ and $g_{\Delta N}$ enter at the one-loop level, the
fractional uncertainty in $g_1$ and $g_{\Delta N}$ will be
parametrically larger than that of
$\left(2\overline{\alpha}-\overline{\beta}\right)/3$ and 
$\overline{\alpha}+\overline{\beta}$ (assuming natural sizes).

\begin{table}[t]
\begin{tabular}{|c|c|c|c|c|c|}
\hline
 Extraction 
& ${1\over 3}\left(2 \overline{\alpha} - \overline{\beta}\right)\ {\rm (l.u.)}$ 
& $ \overline{\alpha} + \overline{\beta}\ {\rm (l.u.)}$ 
& $ g_1$ 
& $ |g_{\Delta N}|$ 
& $\chi^2$/dof 
\\
\cline{1-6}
\cline{1-6}
\ \ LO\ \ ${\cal O}(m_q)$ \ \ & 
$ 0.198\pm 0.093 $
&$ 2.07\pm 0.08$& $ -- $  & $ --$ & $0.56$\\
\cline{1-6} 
\cline{1-6} 
\ \ NLO\ \ ${\cal O}(m_q^{3/2})$\ \ & 
 $\quad 0.229\pm 0.058 \quad$
& $\quad 3.4\pm 1.1\quad$ & $\quad -0.10\pm 0.35\quad$  & $\quad 0.60\pm 0.66\quad$ & $0.21$\\
\cline{1-6} 
\cline{1-6} 
\end{tabular}
\caption{Parameter Table.  The values of the parameters in the
partially-quenched chiral Lagrangian as determined by a
$\chi^2$-minimization fit of the theoretical proton mass differences
given in Appendix~\protect\ref{app:pqmass}, to the lattice data given
in Table~\protect\ref{table:pq}.  
The isospin-conserving combination of counterterms,
$\overline{\alpha}+\overline{\beta}$, is renormalization-scale
dependent.  We have renormalized at $\mu=1~{\rm GeV}$.  }
\label{table:params}
\end{table}
At this order there are contributions from loops diagrams involving
nucleons and diagrams involving $\Delta$'s.  The axial coupling
between the nucleons and the pions is taken from the chiral
perturbation fit to the recent LHPC lattice
calculation~\cite{Edwards:2005ym}.  We use a value of $g_A$ that is
the average of $g_A=1.25$, corresponding to the QCD point on the $b
m_l=0.007$ lattices, and $g_A=1.24$, corresponding to the QCD point on
the $b m_l=0.010$ lattices.  Similarly, for the pion decay constant,
we use a value of $f_\pi=149.8$, which is the average of
$f_\pi=147.8~{\rm MeV}$ corresponding to the QCD point on the $b
m_l=0.007$ lattices, and $f_\pi=151.8~{\rm MeV}$ corresponding to the
QCD point on the $b m_l=0.010$ lattices.  We fit $g_{\Delta N}$ to the
data, and find $|g_{\Delta N}^{\rm fit}| = 0.60\pm 0.66$.  This value
is slightly smaller than the $SU(6)$ value of $g_{\Delta N} = -{6\over
5} g_A$~\footnote{ A three-parameter fit with $|g_{\Delta N}|=1.8$
(the value obtained at tree-level from $\Delta\rightarrow N\pi$)
yields $M_n-M_p\big|^{d-u} = 2.06\pm 0.99\pm 0.38~{\rm MeV}$ while
fitting with nucleon loops alone ($g_{\Delta N}=0$) gives
$M_n-M_p\big|^{d-u} = 2.29\pm 0.53 \pm 0.43~{\rm MeV}$.}.  The mass
difference between the $\Delta$ and nucleon must be input into the
extraction.  Ideally, this would also be fit to the data, but we do
not have precise enough data to permit such a fit.  As the
neutron-proton mass difference is highly insensitive to this
quantity, we use the experimental value $M_\Delta-M_N = 293~{\rm
MeV}$. At one-loop order, ${\cal O}(m_q^{3/2})$, we find
\begin{eqnarray}
M_n-M_p\big|^{d-u} & = & 2.26\pm 0.57 \pm 0.42 \pm 0.10~{\rm MeV}
\ \ \ ,
\label{eq:looplevelnumbers}
\end{eqnarray}
where the last error is an estimate of the systematic error due to
truncation of the chiral expansion. It is reassuring that the
predicted neutron-proton mass difference is relatively insensitive to
the order in the chiral expansion, as shown in
Table~\ref{table:properties}. Both the tree-level and the one-loop
extraction of the neutron-proton mass differences are consistent with
the ``experimental'' value of $M_n-M_p\big|^{d-u} = 2.05\pm 0.30~{\rm
MeV}$.
\begin{table}[t]
\begin{tabular}{|c|c|}
\hline
\  Extraction\  
&\ \   $M_n-M_p\big|^{d-u}$\  (MeV) \  at\ $ m_\pi^{\rm phys.}$\ \ 
\\
\cline{1-2}
\cline{1-2}
\ LO\ \ ${\cal O}(m_q)$ & $1.96\pm 0.92 \pm 0.37$\   \\
\cline{1-2} 
\cline{1-2} 
\ NLO\ \ ${\cal O}(m_q^{3/2})$ \  & $2.26 \pm 0.57 \pm 0.42$  
\\
\cline{1-2} 
\cline{1-2} 
\end{tabular}
\caption{
The neutron-proton mass-splitting 
at the physical value of the pion mass,
$m_\pi^{\rm phys.}= 140~{\rm MeV}$,
extracted from this partially-quenched lattice calculation, using the 
parameters shown in Table~\protect\ref{table:params}.
The lattice spacing used to convert between lattice units and physical units
is $b=0.125~{\rm fm}$.
The first error is statistical while
the second error is due to the uncertainty in the ratio of quark masses,
$m_u/m_d$, in the MILC calculation~\protect\cite{Aubin:2004fs}.
}
\label{table:properties}
\end{table}

An interesting observation can be made by comparing the proton mass
differences on the two different lattice sets, as shown in
Table~\ref{table:pq} and displayed in fig.~\ref{fig:pqdata}.  Within
errors, the magnitude of the mass differences are independent of the
value of the sea-quark mass.  This is consistent with the
leading-order chiral expansions given eq.~(\ref{eq:protmass}) and in
Appendix~\ref{app:pqmass}.  Higher-order contributions to these mass
differences in the chiral expansion, which give rise to deviations
from these equalities, will be become more visible with increased
statistics.

There will be finite lattice spacing contributions to the parameters
that we have extracted in this work.  The recent developments in the
inclusion of finite-lattice spacing effects in mixed-action theories
in $\chi$PT allow us to determine where such corrections enter and to
estimate how big the corrections should be.  The lattice spacing is
introduced into the mixed-action theory by extending the
$SU(2)_L\otimes SU(2)_R$ lie-algebra to a graded lie-algebra that makes the distinction between sea
and valence quarks explicit.  The lattice spacing is incorporated by a
spurion field with the appropriate transformation properties under the
graded group, e.g. see
Ref.~\cite{Bar:2002nr,Bar:2003mh,Bar:2005tu,Tiburzi:2005is}.  There is
a leading-order contribution at ${\cal O}(a^2\ m_q^0)$ to the nucleon
mass (where we are assuming that the exponentially-suppressed
contribution at ${\cal O}(a\ m_q^0)$ from the finite $L_5$ is
numerically insignificant).  However, such terms do not contribute to
the mass differences between the proton states that we have used to
extract the parameters.  Finite lattice spacing contributions to the
nucleon mass that depend upon the light-quark masses start at ${\cal
O}(a^2\ m_q)$.  Therefore, we expect the finite lattice spacing
corrections to our results to be parametrically suppressed and small.
In contrast, the finite lattice spacing contribution to the nucleon
mass itself is expected to be roughly the same size as the
contribution from the $\sigma$-term, rendering an extraction of the
nucleon $\sigma$-term (e.g. see Ref.~\cite{Leinweber:2003dg}) somewhat
unclean if the physical value of the nucleon mass is used in the
extraction. Finite-volume effects in the baryon mass splittings are
estimated to be negligible~\cite{Beane:2004tw}.

\section{Conclusions}
\label{sec:resdisc}

\noindent 
In this work we have performed the first lattice calculation of the
neutron-proton mass-difference arising from the difference between the
mass of the up and down quarks, and find $M_n-M_p\big|^{d-u} = 2.26\pm
0.57 \pm 0.42 \pm 0.10~{\rm MeV}$.  This value is consistent with the
number $2.05\pm 0.30~{\rm MeV}$ based upon the experimentally-measured
mass-difference and the best estimate of the electromagnetic
contribution.  It is clear that further lattice calculations are
warranted in order to make a precise prediction for this quantity,
which will then enable a precise determination of the electromagnetic
contribution to this mass-difference~\footnote{For an attempt to
determine the electromagnetic contribution to the neutron-proton mass
difference using quenched lattice QCD, see
Ref.~\cite{Duncan:1996be}.}.

The neutron-proton mass difference is but one of the manifestations of
charge-symmetry breaking that has been a focus of both theoretical and
experimental investigations for many years~\cite{Miller:2006tv}.  With
the recent progress in extracting the nucleon-nucleon scattering
amplitudes from fully-dynamical lattice QCD~\cite{Beane:2006mx}, one
can imagine using partially-quenched calculations to extract the
charge-symmetry breaking contribution to nucleon-nucleon scattering,
and other processes, in future investigations.

\acknowledgments

\noindent 
This work was performed under the auspices of SciDAC.  We acknowledge
useful discussions with Jerry Miller on charge-symmetry breaking.  We
thank R.~Edwards for help with the QDP++/Chroma programming
environment~\cite{Edwards:2004sx} with which the calculations
discussed here were performed. We are also indebted to the MILC and
the LHP collaborations for use of some of their configurations and
propagators, respectively.  This work was supported in part by the
U.S.~Dept.~of Energy under Grants No.~DE-FG03-97ER4014 (MJS) and
No.~DF-FC02-94ER40818 (KO), the National Science Foundation under
grant No.~PHY-0400231 (SRB) and by DOE through contract
DE-AC05-84ER40150, under which the Southeastern Universities Research
Association (SURA) operates the Thomas Jefferson National Accelerator
Facility (KO,SRB).

\vfill\eject

\appendix

\section{Nucleon Mass in PQHB$\chi$PT}
\label{app:pqmass}

\noindent In this appendix we give the expression for a proton
composed of valence quarks $V_1V_1V_1,\ V_1V_1V_2, \ V_2V_2V_1$ and
$V_2V_2V_2$ on an isospin-symmetric sea composed of two light quarks
$V_1, V_1$.  The proton masses are, starting with the QCD point,
\begin{eqnarray}
& & M_p(V_1,V_1,V_1;V_1) \ =\  M_0 \ +\ 
\left(\overline{\alpha} \ +\  \overline{\beta} \ +\ 2 \
  \overline{\sigma}\right)\  m^2_{V_1,V_1;V_1}
\nonumber\\
& & 
\qquad
\ -\ 
{3 g_A^2\over 16\pi f^2} m^3_{V_1,V_1;V_1} 
\ -\ 
{g_{\Delta N}^2\over 6\pi^2 f^2} F_{V1,V1;V1} \ \ ;
\nonumber\\
& & 
\nonumber\\
& & M_p(V_1,V_1,V_2;V_1)\ =\  M_0 \ +\  2 \ \overline{\sigma}\
m^2_{V_1,V_1;V_1}    
\ +\
{1\over 6}\left(5\overline{\alpha} \ +\  2\overline{\beta}\right)\ m^2_{V_1,V_1;V_1}
\nonumber\\
&& 
\qquad
\ +\
{1\over 6}\left(\overline{\alpha} \ +\  4\overline{\beta}\right)\  m^2_{V_2,V_2;V_1}
\nonumber\\
& & 
\qquad
\ -\ 
{g_A^2\over 24\pi f^2} \left( {7\over 2} m^3_{V_1,V_1;V_1} \ +\
  m^3_{V_1,V_2;V_1} \right)
\nonumber\\
& & 
\qquad
\ -\ 
{g_A g_1\over 24\pi f^2} \left( {5\over 2} m^3_{V_1,V_1;V_1}  \ -\
  m^3_{V_1,V_2;V_1} 
\ -\ {3\over 2} m^3_{V_2,V_2;V_1}
\right)
\nonumber\\
& & 
\qquad
\ -\ 
{g_1^2\over 384\pi f^2} \left( 14 m^3_{V_1,V_1;V_1} 
\ +\  4 m^3_{V_1,V_2;V_1} 
\ -\ 27 m^3_{V_2,V_2;V_1}
\ +\ 9 m^2_{V_1,V_1;V_1} m_{V_2,V_2;V_1}
\right)
\nonumber\\
& & 
\qquad
\ -\ 
{g_{\Delta N}^2\over 72\pi^2 f^2} 
\left(\ 2 F_{V_1,V_1;V_1}  + 9 F_{V_1,V_2;V_1} + F_{V_2,V_2;V_1} + 
m^2_{V_1,V_1;V_1} S_{V_2,V_2;V_1}
 - 
m^2_{V_2,V_2;V_1} S_{V_2,V_2;V_1}
\right);
\nonumber\\
& & 
\nonumber\\
& & M_p(V_2,V_2,V_1;V_1) \ = \ M_0 \ +\  2 \ \overline{\sigma}\  m^2_{V_1,V_1;V_1}    \ +\
{1\over 6}\left(5\overline{\alpha} \ +\  2\overline{\beta}\right)\  m^2_{V_2,V_2;V_1}
\nonumber\\
&& 
\qquad
\ +\
{1\over 6}\left(\overline{\alpha} \ +\  4\overline{\beta}\right)\  m^2_{V_1,V_1;V_1}
\nonumber\\
&& 
\qquad
\ - \
{g_A^2\over 96\pi f^2} \left( 20 m^3_{V_1,V_2;V_1} \ +\
 9 m^2_{V_1,V_1;V_1} m_{V_2,V_2;V_1}
\ -\ 11 m^3_{V_2,V_2;V_1}
\right)
\nonumber\\
& & 
\qquad
\ -\ 
{g_A g_1\over 96\pi f^2} \left( 4 m^3_{V_1,V_2;V_1}  \ +\
  9 m^2_{V_1,V_1;V_1} m_{V_2,V_2;V_1}
\ -\ 13 m^3_{V_2,V_2;V_1}
\right)
\nonumber\\
& & 
\qquad
\ -\ 
{g_1^2\over 384\pi f^2} \left( 
18 m^3_{V_1,V_1;V_1} 
\ -\  4 m^3_{V_1,V_2;V_1} 
\ -\ 23 m^3_{V_2,V_2;V_1}
\ +\ 9 m^2_{V_1,V_1;V_1} m_{V_2,V_2;V_1}
\right)
\nonumber\\
& & 
\qquad
\ -\ 
{g_{\Delta N}^2\over 72\pi^2 f^2} 
\left(\ 
3 F_{V_1,V_1;V_1}  + 7 F_{V_1,V_2;V_1} + 2 F_{V_2,V_2;V_1} + 
m^2_{V_1,V_1;V_1} S_{V_2,V_2;V_1}
 - 
m^2_{V_2,V_2;V_1} S_{V_2,V_2;V_1}
\right) ;
\nonumber\\
& & 
\nonumber\\
& & 
M_p(V_2,V_2,V_2;V_1) \ =\ M_0 \ +\  2 \ \overline{\sigma}\  m^2_{V_1,V_1;V_1}    \ +\
\left(\overline{\alpha} \ +\  \overline{\beta}\right)\  m^2_{V_2,V_2;V_1}
\nonumber\\
& & 
\qquad
\ - \
{g_A^2\over 96\pi f^2} 
\left( 16 m^3_{V_1,V_2;V_1} \ +\
 9 m^2_{V_1,V_1;V_1} m_{V_2,V_2;V_1}
\ -\ 7 m^3_{V_2,V_2;V_1}
\right)
\nonumber\\
& & 
\qquad
\ -\ 
{g_A g_1\over 48\pi f^2} \left( 
4 m^3_{V_1,V_2;V_1}  \ +\
  9 m^2_{V_1,V_1;V_1} m_{V_2,V_2;V_1}
\ -\ 13 m^3_{V_2,V_2;V_1}
\right)
\nonumber\\
& & 
\qquad
\ -\ 
{g_1^2\over 96\pi f^2} \left( 
10 m^3_{V_1,V_2;V_1} 
\ +\ 9 m^2_{V_1,V_1;V_1} m_{V_2,V_2;V_1}
\ -\ 19 m^3_{V_2,V_2;V_1} 
\right)
\nonumber\\
& & 
\qquad
\ -\ 
{g_{\Delta N}^2\over 12\pi^2 f^2} 
\left(\ 
F_{V_1,V_2;V_1} \ +\  F_{V_2,V_2;V_1}
\right)
\ \ ,
\label{eq:pqmassNPI}
\end{eqnarray}
where we have used the leading-order relation between the quark masses,
and the function $S_\pi = S(m_\pi,\Delta,\mu)$ is
\begin{eqnarray}
S(m,\Delta,\mu) & = & 
\sqrt{\Delta^2-m^2} 
\log\left({\Delta -\sqrt{\Delta^2-m^2+i\epsilon}\over
\Delta +\sqrt{\Delta^2-m^2+i\epsilon}}\right)
-\Delta \left( \log\left({m^2\over\mu^2}\right) + {1\over 3} \right)
\ \ \ .
\label{eq:massfunS}
\end{eqnarray}
The function $F$ is defined in the text.

The meson masses at leading order in the chiral expansion, starting with the
QCD point,
are
\begin{eqnarray}
m_\pi^2(V_1,V_1;V_1)\ =\ 
m^2_{V_1,V_1;V_1} & = & 2\ \lambda\  m_{V_1}
\nonumber\\
m_\pi^2(V_1,V_2;V_1)\ =\ 
m^2_{V_1,V_2;V_1} & = & \lambda\ \left( m_{V_1} +  m_{V_2} \right)
\nonumber\\
m_\pi^2(V_2,V_2;V_1)\ =\ 
m^2_{V_2,V_2;V_1} & = & 2 \ \lambda\  m_{V_2}
\ \ ,
\label{eq:pqmasspi}
\end{eqnarray}
where $\lambda$ is a strong interaction mass-scale, and the semicolon separates
valence and sea quarks.

\end{document}